%% file: main.tex
\documentclass{article}

\usepackage[nonatbib,final]{neurips_2019}

\usepackage[utf8]{inputenc} 
\usepackage[T1]{fontenc}    
\usepackage{hyperref}       
\usepackage{url}            
\usepackage{booktabs}       
\usepackage{amsfonts}       
\usepackage{nicefrac}       
\usepackage{microtype}      

\usepackage[capitalise]{cleveref}

\usepackage{graphicx}
\usepackage{color}

\usepackage[style=numeric,sorting=none,minnames=10,maxnames=15]{biblatex}
\bibliography{refs}

\usepackage{amsmath} 
\usepackage{amsfonts} 
\usepackage{todonotes}

\usepackage{custom_macros}

\renewcommand{\top}{\mathsf{T}}

\newcommand{\bLam}{\boldsymbol{\Lambda}}
\newcommand{\bell}{\boldsymbol{\ell}}
\newcommand{\bzero}{\boldsymbol{0}}

\usepackage{caption}
\title{Universality and individuality in neural dynamics across large populations of recurrent networks}

\author{%
  Niru Maheswaranathan\thanks{Equal contribution.} \\
  Google Brain, Google Inc. \\
  Mountain View, CA \\
  \texttt{nirum@google.com } \\
  \And
  Alex H. Williams$^{*}$ \\
  Stanford University \\
  Stanford, CA \\
  \texttt{ahwillia@stanford.edu} \\
  \And
  Matthew D. Golub \\
  Stanford University \\
  Stanford, CA \\
  \texttt{mgolub@stanford.edu} \\
  \And
  Surya Ganguli \\
  Stanford University and Google Brain \\
  Stanford, CA and Mountain View, CA \\
  \texttt{sganguli@stanford.edu} \\
  \And
  David Sussillo\thanks{Corresponding author.} \\
  Google Brain, Google Inc. \\
  Mountain View, CA \\
  \texttt{sussillo@google.com } \\
}

\begin{document}

\maketitle
\input{00-abstract.tex}

\input{01-introduction.tex}

\input{02-methods.tex}
\input{03-results.tex}
\input{04-related_work.tex}
\input{05-discussion.tex}
\subsubsection*{Acknowledgments}
The authors would like to thank Jeffrey Pennington, Maithra Raghu, Jascha Sohl-Dickstein, and Larry Abbott for helpful feedback and discussions. MDG was supported by the Stanford Neurosciences Institute, the Office of Naval Research Grant \#N00014-18-1-2158.

\printbibliography

\newpage
\appendix
\input{06-appendix.tex}

\end{document}

%% file: 00-abstract.tex
\begin{abstract}
Task-based modeling with recurrent neural networks (RNNs) has emerged as a popular way to infer the computational function of different brain regions. These models are quantitatively assessed by comparing the low-dimensional neural representations of the model with the brain, for example using canonical correlation analysis (CCA). However, the nature of the detailed neurobiological inferences one can draw from such efforts remains elusive. For example, to what extent does training neural networks to solve common tasks uniquely determine the network dynamics, independent of modeling architectural choices?  Or alternatively, are the learned dynamics highly sensitive to different model choices?  Knowing the answer to these questions has strong implications for whether and how we should use task-based RNN modeling to understand brain dynamics. To address these foundational questions, we study populations of thousands of networks, with commonly used RNN architectures, trained to solve neuroscientifically motivated tasks and characterize their nonlinear dynamics. We find the geometry of the RNN representations can be highly sensitive to different network architectures, yielding a cautionary tale for measures of similarity that rely on representational geometry, such as CCA. Moreover, we find that while the geometry of neural dynamics can vary greatly across architectures, the underlying computational scaffold---the topological structure of fixed points, transitions between them, limit cycles, and linearized dynamics---often appears universal across all architectures.


\end{abstract}

%% file: 01-introduction.tex
\section{Introduction}

The computational neuroscience community is increasingly relying on deep learning both to directly model large-scale neural recordings \cite{McIntosh2016, maheswaranathan2018deep, Pandarinath2018} as well to train neural networks on computational tasks and compare the internal dynamics of such trained networks to measured neural recordings \cite{Mante2013, Kell2018, Rajalingham2018, cueva2018emergence, banino2018vector, recanatesi2019}. For example, several recent studies have reported similarities between the internal representations of biological and artificial networks \cite{Kell2018, Yamins2014, Khaligh2014, sussillo2015neural, Remington2018, wang2018flexible, Barrett2019, Orhan2019}.  These representational similarities are quite striking since artificial neural networks clearly differ in many ways from their much more biophysically complex natural counterparts.  How then, should we scientifically interpret the striking representational similarity of biological and artificial networks, despite their vast disparity in biophysical and architectural mechanisms?

A fundamental impediment to achieving any such clear scientific interpretation lies in the fact that infinitely many model networks may be consistent with any particular computational task or neural recording.  Indeed, many modern applications of deep learning utilize a wide variety of recurrent neural network (RNN) architectures \cite{Hochreiter1997, Cho2014, Wisdom2016, Collins2017}, initialization strategies \cite{Le2015} and regularization terms \cite{Pascanu2013, Merity2018}. Moreover, new architectures continually emerge through large-scale automated searches \cite{Zoph2017, Pham2018, Chen2018}. This dizzying set of modelling degrees of freedom in deep learning raises fundamental questions about how the degree of match between dynamical properties of biological and artificial networks varies across different modelling choices used to generate RNNs. 

For example, do certain properties of RNN dynamics vary widely across individual architectures?  If so, then a high degree of match between these properties measured in both an artificial RNN and a biological circuit might yield insights into the architecture underlying the biological circuit's dynamics, as well as rule out other potential architectures.  Alternatively, are other properties of RNN dynamics {\it universal} across many architectural classes and other modelling degrees of freedom?  If so, such properties are interesting neural invariants determined primarily by the task, and we should naturally expect them to recur not only across diverse classes of artificial RNNs, but also in relevant brain circuits that solve the same task. The existence of such universal properties would then provide a satisfying explanation of certain aspects of the match in internal representations between biological and artificial RNNs, despite many disparities in their underlying mechanisms.

Interestingly, such universal properties can also break the vast design space of RNNs into different {\it universality classes}, with these universal dynamical properties being constant within classes, and varying only between classes.  This offers the possibility of theoretically calculating or understanding such universal properties by analyzing the simplest network within each universality class\footnote{This situation is akin to that in equilibrium statistical mechanics in which physical materials as disparate as water and ferromagnets have {\it identical} critical exponents at second order phase transitions, by virtue of the fact that they fall within the same universality class \cite{Stanley1971-go}. Moreover, these universal critical exponents can be computed theoretically in the simplest model within this class: the Ising model.}. Thus a foundational question in the theory of RNNs, as well as in their application to neuroscientific modelling, lies in ascertaining which aspects of RNN dynamics vary across different architectural choices, and which aspects---if any---are universal across such choices. 


Theoretical clarity on the nature of individuality and universality in nonlinear RNN dynamics is largely lacking\footnote{Although Feigenbaum's analysis \cite{feigenbaum2017universal} of period doubling in certain 1D maps might be viewed as an analysis of 1D RNNs.}, with some exceptions \cite{rivkind2017local,mastrogiuseppe2018linking,mastrogiuseppe2019geometrical,Saxe2019-eg}. Therefore, with the above neuroscientific and theoretical motivations in mind, we initiate an extensive numerical study of the variations in RNN dynamics across thousands of RNNs with varying modelling choices.  We focus on canonical neuroscientifically motivated tasks that exemplify basic elements of neural computation, including the storage and maintenance of multiple discrete memories, the production of oscillatory motor-like dynamics, and contextual integration in the face of noisy evidence \cite{Sussillo2013, Mante2013}.


To compare internal representations across networks, we focused on comparing the geometry of neural dynamics using common network similarity measures such as singular vector canonical correlation analysis (SVCCA) \cite{Raghu2017} and centered kernel alignment (CKA) \cite{kornblith2019similarity}. We also used tools from dynamical systems analysis to extract more topological aspects of neural dynamics, including fixed points, limit cycles, and transition pathways between them, as well as the linearized dynamics around fixed points \cite{Sussillo2013}.  We focused on these approaches because comparisons between artificial and biological network dynamics at the level of geometry, and topology and linearized dynamics, are often employed in computational neuroscience.


Using these tools, we find that different RNN architectures trained on the same task exhibit both universal and individualistic dynamical properties.  In particular, we find that the geometry of neural representations varies considerably across RNNs with different nonlinearities.  We also find surprising dissociations between dynamical similarity and functional similarity, whereby trained and untrained architectures of a given type can be more similar to each other than trained architectures of different types.  This yields a cautionary tale for using SVCCA or CKA to compare neural geometry, as these similarity metrics may be more sensitive to particular modeling choices than to overall task performance.  Finally, we find considerably more universality across architectures in the topological structure of fixed points, limit cycles, and specific properties of the linearized dynamics about fixed points.  Thus overall, our numerical study provides a much needed foundation for understanding universality and individuality in network dynamics across various RNN models, a question that is both of intrinsic theoretical interest, and of importance in neuroscientific applications.

%% file: 02-methods.tex
\section{Methods}

\subsection{Model Architectures and Training Procedure}

We define an RNN by an update rule, $\bh_t = F (\bh_{t-1}, \bx_t)$, where $F$ denotes some nonlinear function of the network state vector $\bh_{t-1} \in \mathbb{R}^N$ and the network input $\bx_t \in \mathbb{R}^M$.
Here, $t$ is an integer index denoting discrete time steps.
Given an initial state, $\bh_0$, and a stream of $T$ inputs, $\bx_1$, $\bx_2$, $\hdots$, $\bx_T$, the RNN states are recursively computed, $\bh_1$, $\bh_2$, $\hdots$, $\bh_T$.
The model predictions are based on a linear readout of these state vector representations of the input stream. We studied 4 RNN architectures, the vanilla RNN (Vanilla), the Update-Gate RNN (UGRNN; \cite{Collins2017}), the Gated Recurrent Unit (GRU; \cite{Cho2014}), and the Long-Short-Term-Memory (LSTM; \cite{Hochreiter1997}). The equations for these RNNs can be found in Appendix \ref{appendix:rnneqs}. For each RNN architecture we modified the (non-gate) point-wise activation function to be either rectified linear (relu) or hyperbolic tangent (tanh). The point-wise activation for the gating units is kept as a sigmoid.


We trained networks for every combination of the following parameters: RNN architecture (Vanilla, UGRNN, LSTM, GRU), activation (relu, tanh), number of units/neurons (64, 128, 256), and L2 regularization (1e-5, 1e-4, 1e-3, 1e-2). This yielded $4\times 2\times 3\times 4$ = 96 unique configurations. For each one of these configurations, we performed a separate random hyperparameter search over gradient clipping values \cite{Pascanu2013} (logarithmically spaced from 0.1 to 10) and the learning rate schedule parameters. The learning rate schedule is an exponentially decaying schedule parameterized by the initial rate (with search range from 1e-5 to 0.1), decay rate (0.1 to 0.9), and momentum (0 to 1). All networks were trained using stochastic gradient descent with momentum \cite{polyak1964some,sutskever2013importance} for 20,000 iterations with a batch size of 64. For each network configuration, we selected the best hyperparameters using a validation set. We additionally trained each of these configurations with 30 random seeds, yielding 2,880 total networks for analysis for each task. All networks achieve low error; histograms of the final loss values achieved by all networks are available in Appendix \ref{appendix:losshist}.

\subsection{Tasks}

We used three canonical tasks that have been previously studied in the neuroscience literature:

\paragraph{$K$-bit flip-flop}
Following \cite{Sussillo2013}, RNNs were provided $K$ inputs taking discrete values in $\{-1, 0, +1\}$. The RNN has $K$ outputs, each of which is trained to remember the last non-zero input on its corresponding input. Here we set $K=3$, so e.g. output 2 remembers the last non-zero state of input 2 (+1 or -1), but ignores inputs 1 and 3. We set the number of time steps, $T$, to 100, and the flip probability (the probability of any input flipping on a particular time step) to 5\%.

\paragraph{Frequency-cued sine wave}
Following \cite{Sussillo2013}, RNNs received a static input, $x \sim \text{Uniform}(0, 1)$, and were trained to produce a unit amplitude sine wave, $\sin(2\pi \omega t)$, whose frequency is proportional to the input: $\omega=0.04x + 0.01$. We set $T=500$ and $dt = 0.01$ (5 simulated seconds total).

\paragraph{Context-dependent integration (CDI)}
Following previous work \cite{Mante2013}, RNNs were provided with $K$ static context inputs and $K$ time-varying white noise input streams. On each trial, all but one context input was zero, thus forming a one-hot encoding indicating which noisy input stream of length $T$ should be integrated. The white noise input was sampled from $\mathcal{N}(\mu,1)$ at each time step, with $\mu$ sampled uniformly between -1 and 1 and kept static across time for each trial. RNNs were trained to report the cumulative sum of the cued white-noise input stream across time. Here, we set $K=2$ and $T=30$.



\subsection{Assessing model similarity}

The central questions we examined were: how similar are the representations and dynamics of different RNNs trained on the same task? To address this, we use approaches that highlight different but sometimes overlapping aspects of RNN function: 

\paragraph{SVCCA and CKA to assess representational geometry} We quantified similarity at the level of \textit{representational geometry} \cite{Kriegeskorte2013}. In essence, this means quantifying whether the responses of two RNNs to the same inputs are well-aligned by some kind of linear transformation.

We focused on \textit{singular vector canonical correlations analysis} (SVCCA; \cite{Raghu2017}), which has found traction in both neuroscience \cite{sussillo2015neural} and machine learning communities \cite{deVries2018, Barrett2019}. SVCCA compares representations in two steps. First, each representation is projected onto their top principal components to remove the effect of noisy (low variance) directions. Typically, the number of components is chosen to retain \textasciitilde95\% of the variance in the representation. Then, canonical correlation analysis (CCA) is performed to find a linear transformation that maximally correlates the two representations. This yields $R$ correlation coefficients, $1 \geq \rho_1 \geq \hdots \geq \rho_R \geq 0$, providing a means to compare the two datasets, typically by averaging or summing the coefficients (see Appendix \ref{appendix:svcca} for further details).

In addition to SVCCA, we explored a related metric, \textit{centered kernel alignment} (CKA; \cite{kornblith2019similarity}). CKA is related to SVCCA in that it also suppresses low variance directions, however CKA weights the components proportional to the singular value (as opposed to removing some completely). We found that using SVCCA and CKA yielded similar results for the purposes of determining whether representations cluster by architecture or activation function so we present SVCCA results in the main text but provide a comparison with CKA in Appendix \ref{appendix:ckaresults}.

\paragraph{Fixed point topology to assess computation} An alternative perspective to representational geometry for understanding computation in RNNs is dynamics. We studied RNN dynamics by reducing their nonlinear dynamics to linear approximations. Briefly, this approach starts by optimizing to find the fixed points $\{ \bh^*_1, \bh^*_2, ... \}$ of an RNN such that $\bh_i^* \approx F(\bh_i^*, \bx^*)$. We use the term {\it fixed point} to also include approximate fixed points, which are not truly fixed but are nevertheless very slow on the time scale of the task.

We set the input ($\bx^*$) to be static when finding fixed points.  These inputs can be thought of as specifying different task conditions. In particular, the static command frequency in the sine wave task and the hot-one context signal in the CDI task are examples of such condition specifying inputs.  Note however, that dimensions of $\bx$ that are time-varying are set to $\bzero$ in $\bx^*$. In particular, the dimensions of the input that represent the input pulses in the $3$-bit memory task and the white noise input streams in the CDI task are set to $\bzero$ in $\bx^*$.

Numerical procedures for identifying fixed points are discussed in \cite{Sussillo2013, Golub2018}.
Around each fixed point, the local behavior of the system can be approximated by a reduced system with linear dynamics:
$$\bh_t \approx \bh^* + \bJ(\bh^*, \bx^*) \left(\bh_{t-1} - \bh^*\right),$$
where $\bJ_{ij}(\bh^*,\bx^*) = \frac{\partial F_i(\bh^*, \bx^*)}{\partial h^*_j}$ denotes the Jacobian of the RNN update rule. We studied these linearized systems using the eigenvector decomposition for non-normal matrices (see Appendix \ref{appendix:eigendecomp} for the eigenvector decomposition). In this analysis, both the topology of the fixed points and the linearizations around those fixed points become objects of interest.

\paragraph{Visualizing similarity with multi-dimensional scaling} For each analysis, we computed network similarity between all pairs of network configurations for a given task, yielding a large (dis-)similarity matrix for each task (for example, we show this distance matrix for the flip-flop task in Fig. \ref{fig:flipflop}c). To visualize the structure in these matrices, we used multi-dimensional scaling (MDS) \cite{borg2003modern} to generate a 2D projection which we used for visualization (Fig. \ref{fig:flipflop}d and f, Fig. \ref{fig:sine}c and e, Fig. \ref{fig:cdi}c and d). For visualization purposes, we separate plots colored by RNN architecture (for a fixed nonlinearity, tanh) and nonlinearity (for a fixed architecture, Vanilla).

%% file: 03-results.tex

\begin{figure}
  \centering
  \includegraphics[width=\linewidth]{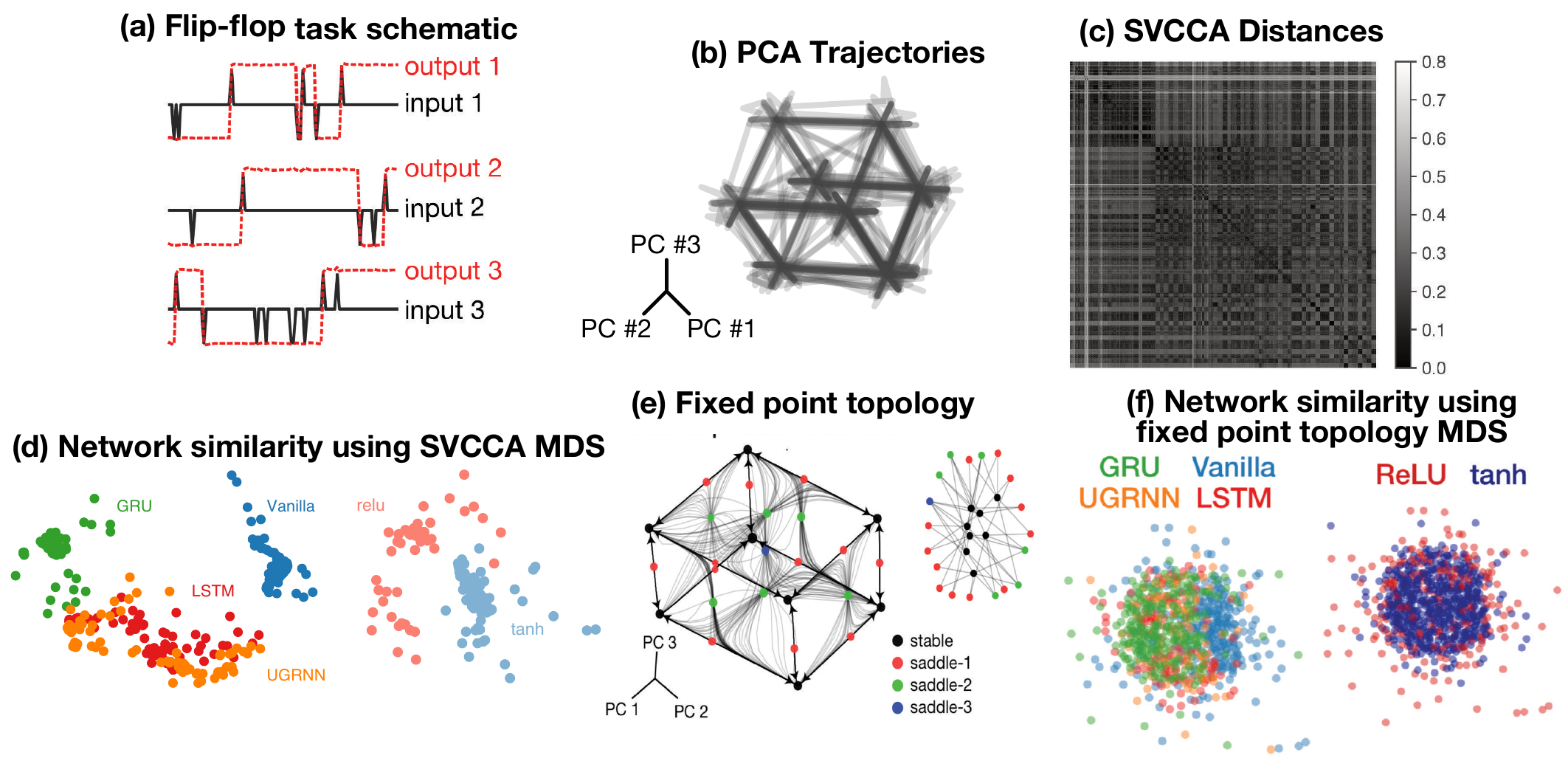}
  \caption{3-bit discrete memory. \textbf{a)} Inputs (black) of -1 or 1 come in at random times while the corresponding output (dashed red) has to remember the last non-zero state of the input (either +1 or -1). \textbf{b)} Example PCA trajectories of dynamics for an example architecture and activation function. \textbf{c)} Dynamics across networks are compared via SVCCA and given a distance (one minus the average correlation coefficient), yielding a network-network distance matrix. \textbf{d)} This distance matrix is used to create a 2D embedding via multidimensional scaling (MDS) of all networks, showing clustering based on RNN architecture (left) and activation function (right). \textbf{e)} Topological analysis of a network using fixed points. First, the fixed points of a network's dynamics are found, and their linear stability is assessed (left, black dots - stable fixed points, red - one unstable dimension, green - 2 unstable dimensions, blue - 3 unstable dimensions.  By studying heteroclinic and homoclinic orbits, the fixed point structure is translated to a graph representation (right).  \textbf{f)} This graph representation is then compared across networks, creating another network-network distance matrix.  The distance matrix is used to embed the network comparisons into 2D space using MDS, showing that the topological representation of a network using fixed point structure is more similar across architectures (left) and activation functions (right) than the geometry of the network is (layout as in 1d).}
  \label{fig:flipflop}
\end{figure}

\section{Results}

The major contributions in this paper are as follows. First, we carefully train and tune large populations of RNNs trained on several canonical tasks relating to discrete memory~\cite{Sussillo2013}, pattern generation~\cite{Sussillo2013}, and analog memory and integration~\cite{Mante2013}. Then, we show that representational geometry is sensitive to model architecture (Figs. 1-3).  Next, we show all RNN architectures, including complex, gated architectures (e.g. LSTM and GRU) converge to qualitatively similar dynamical solutions, as quantified by the topology of fixed points and corresponding linearized dynamics (Figs. 1-3). Finally, we highlight a case where SVCCA is not necessarily indicative of functional similarity (Fig. 4).


\begin{figure}[t!]
  \centering
  \includegraphics[width=\linewidth]{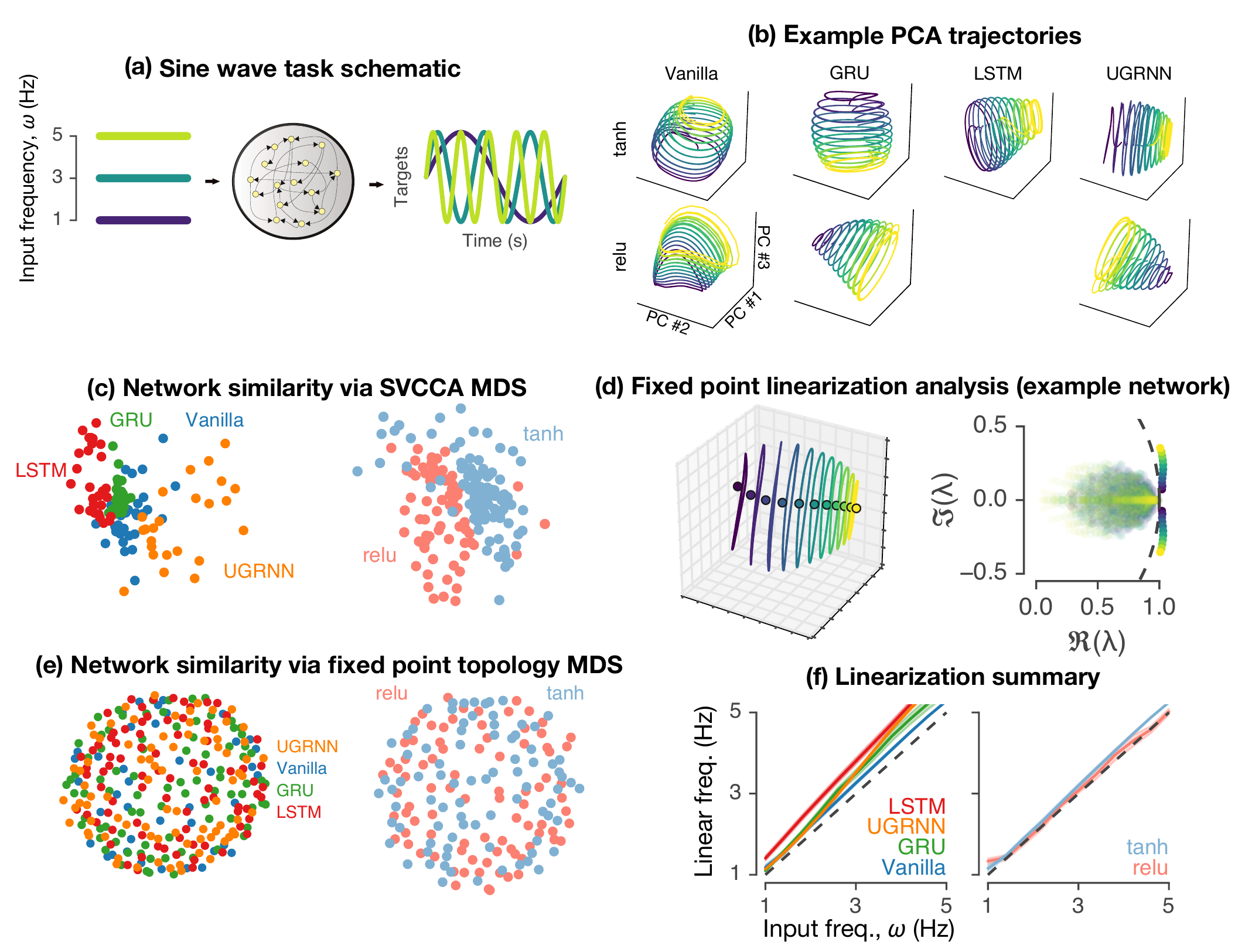}
  \caption{Sine wave generation. \textbf{a)} Schematic showing conversion of static input specifying a command frequency, $\omega$, for the sine wave output $\sin(2\pi\omega t)$. \textbf{b)} PCA plots showing trajectories using many evenly divided command frequencies delivered one at a time (blue: smallest $\omega$, yellow: largest $\omega$). \textbf{c)} MDS plots based on SVCCA network-network distances, layout as in Fig. 1d. \textbf{d)} Left, fixed points (colored circles, with color indicating $\omega$, one fixed point per command frequency) showing a single fixed point in the middle of each oscillatory trajectory. Right, the complex eigenvalues of all the linearized systems, one per fixed point, overlayed on top of each other, with primary oscillatory eigenvalues colored as in panel b. \textbf{e)} MDS network-network distances based on fixed point topology, assessing systematic differences in the topology of the input-dependent fixed points (layout as in Fig. 1d). \textbf{f)} Summary analysis showing the frequency of the oscillatory mode in the linearized system vs. command frequency for different architectures (left) and activations (right). Solid line and shaded patch show the mean $\pm$ standard error over networks trained with different random seeds. Small, though systematic, variations exist in the frequency of each oscillatory mode.}
  \label{fig:sine}
\end{figure}

\subsection{3-bit discrete memory}
We trained RNNs to store and report three discrete binary inputs (\cref{fig:flipflop}a).
In \cref{fig:flipflop}b, we use a simple ``probe input'' consisting of a series of random inputs to highlight the network structure. Across all network architectures the resulting trajectories roughly trace out the corners of a three-dimensional cube. While these example trajectories look qualitatively similar across architectures, SVCCA revealed systematic differences. This is visible in the raw SVCCA distance matrix (\cref{fig:flipflop}c), as well as in low-dimensional linear embeddings achieved by applying multi-dimensional scaling (MDS) (\cref{fig:flipflop}d) created using the SVCCA distance matrix.

To study the dynamics of these networks, we ran an optimization procedure~\cite{Golub2018} to numerically identify fixed points for each trained network (see Methods). A representative network is shown in \cref{fig:flipflop}e (left). The network solves the task by encoding all $2^3$ possible outputs as 8 stable fixed points. Furthermore, there are saddle points with one, two, or three unstable dimensions (see caption), which route the network activity towards the appropriate stable fixed point for a given input.

We devised an automated procedure to quantify the computational logic of the fixed point structure in \cref{fig:flipflop}e that effectively ignored the precise details in the transient dynamics and overall geometry of the 3D cube evident in the PCA trajectories. Specifically, we distilled the dynamical trajectories into a directed graph, with nodes representing fixed points, and weighted edges representing the probability of moving from one fixed point to another when starting the initial state a small distance away from the first fixed point. We did this 100 times for each fixed point, yielding a probability of transitioning from one fixed point to another. As expected, stable fixed points have no outgoing edges, and only have a self-loop. All unstable fixed points had two or more outgoing edges, which are directed at nearby stable fixed points. We constructed a fixed point graph for each network and used the Euclidean distance between the graph connectivity matrices to quantify dis-similarity\footnote{While determining whether two graphs are isomorphic is a challenging problem in general, we circumvented this issue by lexographically ordering the fixed points based on the RNN readout. Networks with different numbers of fixed points than the modal number were discarded (less than 10\% of the population).}. These heteroclinic orbits are shown in \cref{fig:flipflop}e, light black trajectories from one fixed point to another. Using this topological measure of RNN similarity, we find that all architectures converge to very similar solutions as shown by an MDS embedding of the fixed point graph (\cref{fig:flipflop}f).

\subsection{Sine wave generation}
We trained RNNs to convert a static input into a sine wave, e.g. convert the command frequency $\omega$ to $\sin(2\pi\omega t)$ (\cref{fig:sine}a). \cref{fig:sine}b shows low-dimensional trajectories in trained networks across all architectures and nonlinearities (LSTM with ReLU did not train effectively, so we excluded it). Each trajectory is colored by the input frequency.
Furthermore, all trajectories followed a similar pattern: oscillations occur in a roughly 2D subspace (circular trajectories), with separate circles for each frequency input separated by a third dimension. We then performed an analogous series of analyses to those used in the previous task.  In particular, we computed the SVCCA distances (raw distances not shown) and used those to create an embedding of the network activity (\cref{fig:sine}c) as a function of either RNN architecture or activation.  These SVCCA MDS summaries show systematic differences in the representations across both architecture and activation.

Moving to the analysis of dynamics, we found for each input frequency a single input-dependent fixed point (\cref{fig:sine}d, left). We studied the linearized dynamics around each fixed point and found a single pair of imaginary eigenvalues, representing a mildly unstable oscillatory mode whose complex angle aligned well with the input frequency (\cref{fig:sine}d, right).  We compared the frequency of the linear model to the input frequency and found generally good alignment.  We averaged the linear frequency across all networks within architecture or activation and found small, but systematic differences (\cref{fig:sine}f). Embeddings of the topological structure of the input-dependent fixed points did not reveal any structure that systematically varied by architecture or activation (\cref{fig:sine}e).

\begin{figure}[t!]
  \centering
  \includegraphics[width=\linewidth]{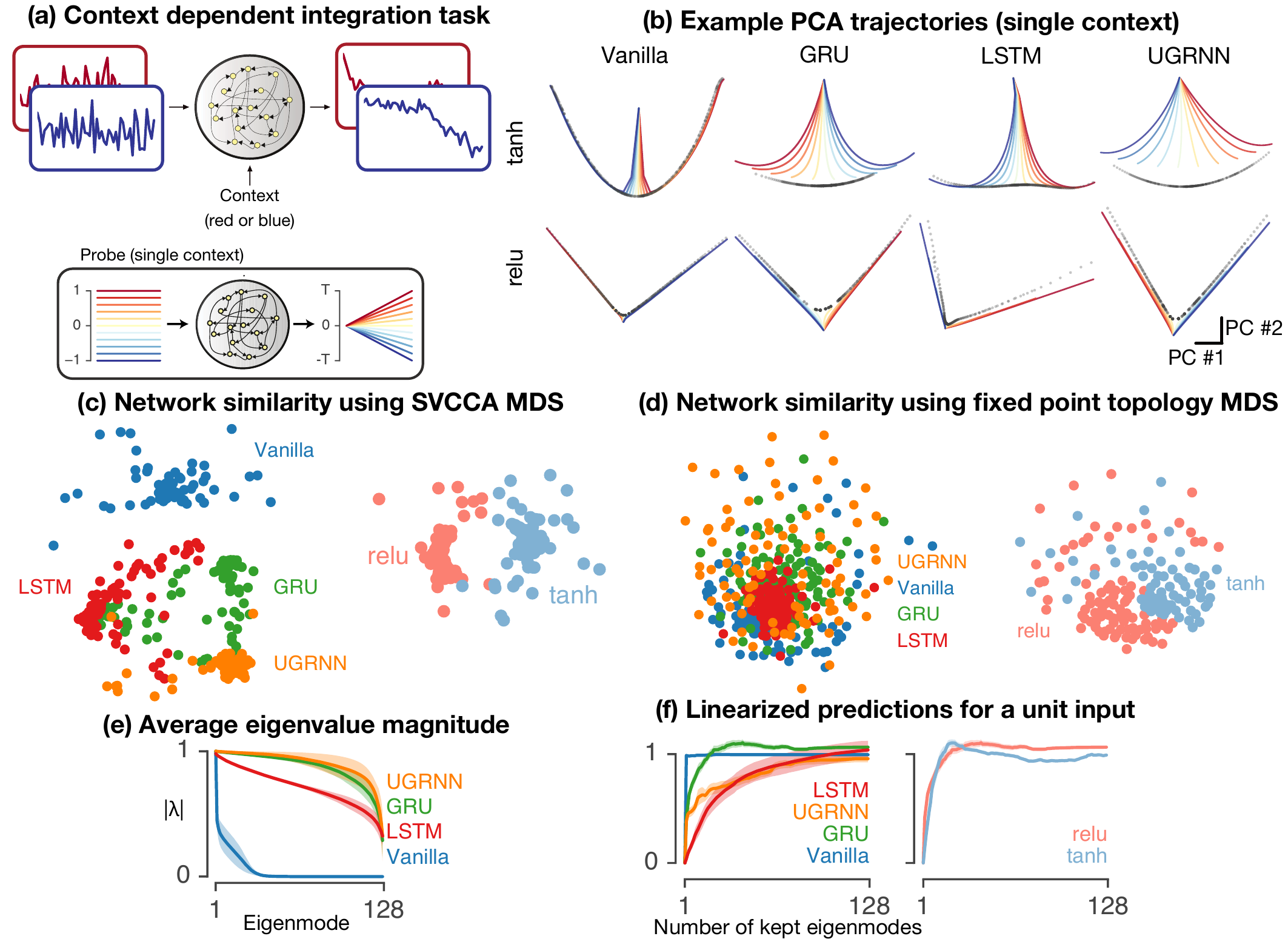}
  \caption{Context-Dependent Integration. \textbf{a)} One of two streams of white-noise input (blue or red) is contextually selected by a one-hot static context input to be integrated as output of the network, while the other is ignored (blue or red). \textbf{b)} The trained networks were studied with probe inputs (panel inset in a), probes from blue to red show probe input). For this and subsequent panels, only one context is shown for clarity. Shown in b are the PCA plots of RNN hidden states when driven by probe inputs (blue to red). The fixed points (black dots) show approximate line attractors for all RNN architectures and nonlinearities. \textbf{c)} MDS embedding of SVCCA network-network distances comparing representations based on architecture (left) and activation (right), layout as in Fig. 1d. \textbf{d)} Using the same method to assess the topology of the fixed points as used in the sine-wave example to study the topology of the input-dependent fixed points, we embedded the network-network distances using the topological structure of the line attractor (colored based on architectures (left) and activation (right), layout as in Fig. 1d). \textbf{e)} Average sorted eigenvalues as a function architecture. Solid line and shaded patch show mean $\pm$ standard error over networks trained with different random seeds. \textbf{f)}  Output of the network when probed with a unit magnitude input using the linearized dynamics, averaged over all fixed points on the line attractor, as a function of architecture and number of linear modes retained. In order to study the the dimensionality of the solution to integration, we systematically removed the modes with smallest eigenvalues one at a time, and recomputed the prediction of the new linear system for the unit magnitude input. These plots indicate that the vanilla RNN (blue) uses a single mode to perform the integration, while the gated architectures distribute this across a larger number of linear modes.}
  \label{fig:cdi}
\end{figure}

\subsection{Context-dependent integration (analog memory)}

We trained an RNN to contextually integrate one of two white noise input streams, while ignoring the other (\cref{fig:cdi}a). We then studied the network representations by delivering a set of probe inputs (\cref{fig:cdi}a). The 3D PCA plots are shown in \cref{fig:cdi}b, showing obvious differences in representational geometry as a function of architecture and activation.  The MDS summary plot of the SVCCA distances of the representations is shown in \cref{fig:cdi}c, again showing systematic clustering as a function of architecture (left) and activation (right). We also analyzed the topology of the fixed points (black dots in \cref{fig:cdi}b) to assess how well the fixed points approximated a line attractor. We quantified this by generating a graph with edges between fixed points that were nearest neighbors. This resulted in a graph for each line attractor in each context, which we then compared using Euclidean distance and embedded in a 2D space using MDS (\cref{fig:cdi}d). The MDS summary plot did not cluster strongly by architecture, but did cluster based on activation.

We then studied the linearized dynamics around each fixed point (\cref{fig:cdi}e,f).  We focused on a single context, and studied how a unit magnitude relevant input (as opposed to the input that should be contextually ignored) was integrated by the linear system around the nearest fixed point. This was previously studied in depth in \cite{Mante2013}.  Here we were interested in differences in integration strategy as a function of architecture.  We found similar results to \cite{Mante2013} for the vanilla RNN, which integrated the input using a single linear mode with an eigenvalue of 1, with input coming in on the associated left eigenvector and represented on the associated right eigenvector. Examination of all linearized dynamics averaged over all fixed points within the context showed that different architectures had a similar strategy, except that the gated architectures had many more eigenvalues near 1 (\cref{fig:cdi}e) and thus used a high-dimensional strategy to accomplish the same goal as the vanilla RNN does in 1 dimension. We further studied the dimensionality by systematically zeroing out eigenvalues from smallest to largest to discover how many linear modes were necessary to integrate a unit magnitude input, compared to the full linear approximation (\cref{fig:cdi}f). These results show that all of the networks and architectures use essentially the same integration strategy, but systematically vary by architecture in terms of the number of modes they employ. To a lesser degree they also vary some in the amount the higher order terms contribute to the solution, as shown by the differences away from an integral of 1 for a unit magnitude input, for the full linearized system with no modes zeroed out (analogous to \cref{fig:sine}f).

\begin{figure}[t!]
  \centering
  \includegraphics[width=0.7\linewidth]{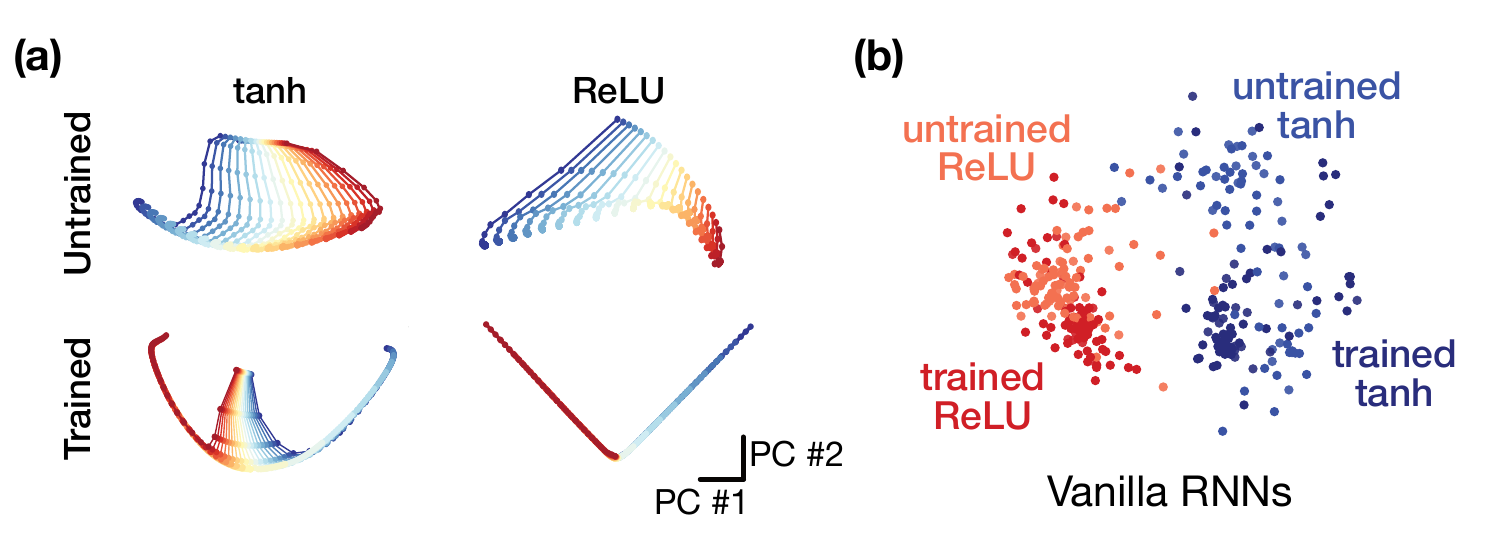}
  \caption{An example where SVCCA yields a stronger correlation between untrained networks and trained networks than between trained networks with different nonlinearities. \textbf{a)} An example (single context shown) of the representation of the probe inputs (blue through red) for four networks: two trained, and two untrained with tanh and ReLU nonlinearities. In this case the untrained tanh and ReLU networks have a higher correlation to the trained tanh network than the trained tanh network does to the trained ReLU network. \textbf{b)} MDS plot of SVCCA-based distances for many trained and untrained networks, showing that trained and untrained relu networks are more similar to each other on average than to tanh networks.}
  \label{fig:cdi_untrained}
\end{figure}

Finally, to highlight the difficulty of using CCA-based techniques to compare representational geometry in simple tasks, we used the inputs of the context-dependent integrator task to drive both trained and untrained vanilla RNNs (\cref{fig:cdi_untrained}). We found that the average canonical correlation between trained and untrained networks can be larger than between trained RNNs with different nonlinearities. The summary MDS plot across many RNNs shows that the two clusters of untrained and trained relu networks are closer together than the two clusters of trained tanh networks \cref{fig:cdi_untrained}b.

%% file: 04-related_work.tex
\section{Related Work}
Researchers are beginning to study both empirically and theoretically how deep networks may show universal properties. For example, \cite{Saxe2019-eg} proved that representational geometry is a universal property amongst all trained deep linear networks that solve a task optimally, with smallest norm weights. Also, \cite{poole2016exponential,raghu2017expressive} studied how expressive capacity increases with network depth and width. Work in RNNs is far more preliminary, though it is well known that RNNs are universal
approximators of dynamical systems \cite{doya1993universality}. More recently, the per-parameter capacity of RNNs was found to be remarkably similar across various RNN architectures \cite{Collins2017}. The authors of \cite{DBLP:journals/corr/abs-1906-01005} studied all the possible topological arrangements of fixed points in a 2D continuous-time GRU, conjecturing that dynamical configurations such as line or ring attractors that require an infinite number of fixed points can only be created in approximation, even in GRUs with more than two dimensions.

Understanding biological neural systems in terms of artificial dynamical systems has a rich tradition \cite{destexhe2009wilson, hopfield1982neural, sompolinsky1988chaos, Seung1996}. Researchers have attempted to understand optimized neural networks with nonlinear dynamical systems techniques \cite{Sussillo2013, barak2013fixed} and to compare those artificial networks to biological circuits \cite{Mante2013, sussillo2015neural, sussillo2014neural, rajan2016recurrent, barak2017recurrent, Remington2018, wang2018flexible}. 

Previous work has studied vanilla RNNs in similar settings \cite{Sussillo2013,Mante2013,yang2019task}, but has not systematically surveyed the variability in network dynamics across commonly used RNN architectures, such as LSTMs \cite{Hochreiter1997} or GRUs \cite{Cho2014}, nor quantified variations in dynamical solutions over architecture and nonlinearity, although \cite{Orhan2019} considers many issues concerning how RNNs may hold memory. Finally, there has been a recent line of work comparing artificial network representations to neural data \cite{McIntosh2016,maheswaranathan2018deep,Pandarinath2018,Yamins2014,Khaligh2014,sussillo2015neural}. Investigators have been studying ways to improve the utility of CCA-based comparison methods \cite{Raghu2017, morcos2018insights}, as well as comparing CCA to other methods \cite{kornblith2019similarity}. 

%% file: 05-discussion.tex

\section{Discussion}
In this work we empirically study aspects of individuality and universality in recurrent networks. We find individuality in that representational geometry of RNNs varies significantly as a function of architecture and activation function (Fig. 1d, 2c, 3c). We also see hints of universality: the fixed point topologies show far less variation across networks than the representations do (Fig. 1f, 2e, 3d). Linear analyses also showed similar solutions, e.g. essentially linear oscillations for the sine wave task (Fig. 2f) and linear integration in the CDI task (Fig. 3f). However, linear analyses also showed variation across architectures in the dimensionality of the solution to integration (Fig. 3e).

While the linear analyses showed common computational strategies across all architectures (such as a slightly unstable oscillation in the linearized system around each fixed point), we did see small systematic differences that clustered by architecture (such as the difference between input frequency and frequency of oscillatory mode in the linearized system). This indicates that another aspect of individuality appears to be the degree to which higher order terms contribute to the total solution.

The fixed point analysis discussed here has one major limitation, namely that the number of fixed points must be the same across networks that are being compared. For the three tasks studied here, we found that the vast majority of trained networks did indeed have the same number of fixed points for each task. However, an important direction for future work is extending the analysis to be more robust with respect to differing numbers of fixed points.

In summary, we hope this empirical study begins a larger effort to characterize methods for comparing RNN dynamics, building a foundation for future connections of biological circuits and artificial neural networks. 



%% file: 06-appendix.tex
\section{RNN Architectures}
\label{appendix:rnneqs}
We examined four RNN architectures that exemplify various degrees of complexity and sophistication. Vanilla RNNs have been historically favored by computational neuroscientists \cite{Mante2013, Remington2018}, while LSTM and GRU networks have been favored by machine learning practitioners due to performance advantages \cite{Collins2017}. However, neuroscientists are beginning to utilize gated RNNs as they progress to studying more complex phenomena \cite{Kanitscheider2017}. Thus, it is of great interest to determine whether similar mechanisms arise across this range of model architectures, or if different models give rise to distinct dynamics and scientific conclusions.
These are architectures are summarized below, with $\bW$ and $\bb$ respectively representing trainable weight matrices and bias parameters.
All other vectors ($\bc, \bg, \br, \bi, \bbf$) represent intermediate quantities; $\sigma()$ represents a pointwise sigmoid nonlinearity; and $f()$ is either the ReLU or tanh nonlineary.

\paragraph{Vanilla RNN}
\begin{equation}
\bh_t = f(\bW^\text{hh} \bh_{t-1} + \bW^\text{hx} \bx_t + \bb^\text{h})
\end{equation}

\paragraph{Update-Gate RNN (UGRNN; \cite{Collins2017})}
\begin{equation}
\begin{aligned}[c]
\bh_t = \bg \cdot \bh_{t-1} + (1 - \bg) \cdot \bc
\end{aligned}
\quad\quad\quad
\begin{aligned}[c]
\bc &= f(\bW^\text{ch} \bh_{t-1} + \bW^\text{cx} \bx_t + \bb^\text{c}) \\
\bg &= \sigma(\bW^\text{gh} \bh_{t-1} + \bW^\text{gx} \bx_t + \bb^\text{g} + b^{fg})
\end{aligned}
\end{equation}

\paragraph{Gated Recurrent Unit (GRU; \cite{Cho2014})}
\begin{equation}
\begin{aligned}[c]
\bh_t = \bg \cdot \bh_{t-1} + (1 - \bg) \cdot \bc
\end{aligned}
\quad\quad\quad
\begin{aligned}[c]
\bc &= f(\bW^\text{ch} (\br \cdot \bh_{t-1}) + \bW^\text{cx} \bx_t + \bb^\text{c}) \\
\bg &= \sigma(\bW^\text{gh} \bh_{t-1} + \bW^\text{gx} \bx_t + \bb^\text{g} + b^{fg}) \\
\br &= \sigma(\bW^\text{rh} \bh_{t-1} + \bW^\text{rx} \bx_t + \bb^\text{r})
\end{aligned}
\end{equation}

\paragraph{Long-Short-Term-Memory (LSTM; \cite{Hochreiter1997})}
\begin{equation}
\hspace{3em}
\begin{aligned}[c]
\bh_t =
\begin{bmatrix}
~\\[-1.5ex]
~\bc_t~ \\[1.2ex]
~\widetilde{\bh}_t~\\[-1.5ex]
~
\end{bmatrix}
\end{aligned}
\hspace{7em}
\begin{aligned}[c]
\widetilde{\bh}_t &= f(\bc_t) \cdot \sigma(\bW^\text{hh} \bh + \bW^\text{hx} \bx + \bb^\text{h})\\
\bc_t &= \bbf_t \cdot \bc_{t-1} + \bi \cdot \sigma(\bW^\text{ch} \widetilde{\bh}_{t-1} + \bW^\text{cx} \bx + \bb^\text{c}) \\
\bi &= \sigma(\bW^\text{ih} \bh + \bW^\text{ix} \bx + \bb^\text{i}) \\
\bbf &= \sigma(\bW^\text{fh} \bh + \bW^\text{fx} \bx + \bb^\text{f} + b^{fg})
\end{aligned}
\end{equation}

\section{Non-normal linear dynamical systems analysis}
\label{appendix:eigendecomp}
We studied the linearized systems, e.g. $\frac{\partial F(\bh^*, \bx^*)_i}{\partial h^*_j}$ using the eigenvector decomposition for non-normal matrices, dropping the dependence on $\bh^*$ and $\bx^*$ for clarity \begin{equation}
\label{eq:impulse-eigdecomp}
\bJ = \bR \bLambda \bL  = \sum_{a=1}^N \lambda_a \br_a \bell_a^\top  ,
\end{equation}
where $\bL = \bR^{-1}$, the columns of $\bR$ (denoted $\br_a$) contain the \textit{right eigenvectors} of $\bJ^\text{rec}$, the rows of $\bL$ (denoted $\bell^\top_a$) contain the \textit{left eigenvectors} of $\bJ^\text{rec}$, and $\bLam$ is a diagonal matrix containing complex-valued eigenvalues, $\lambda_1 > \lambda_2 > \hdots > \lambda_N$, which are sorted based on their magnitude. Note in particular there is no requirement that $\bR^\top\bR = \mathbf{I}$, leading to potentially sophisticated locally linear dynamics.

\section{Network performance}
\label{appendix:losshist}
Below, we include a plot of the final performance of all networks retained for analysis in this paper. All networks achieve low error for their respective tasks.

\begin{figure}[ht!]
    \centering
    \includegraphics[width=\linewidth]{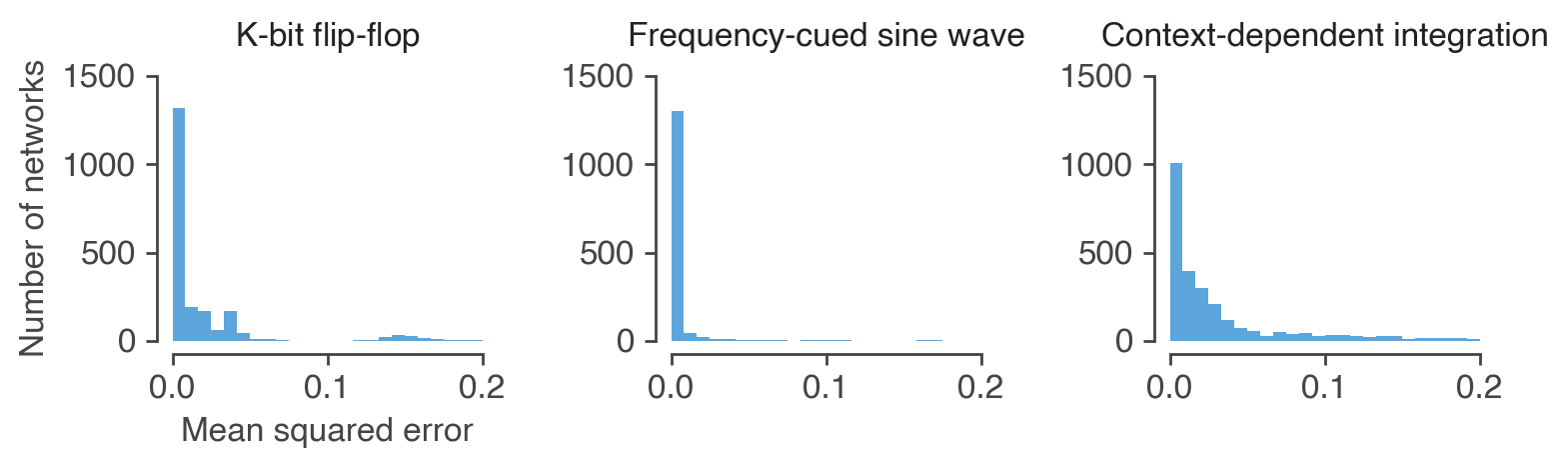}
    \caption{Histogram of final performance (mean squared error) across all networks used for the analyses in this paper.}
    \label{fig:losshist}
\end{figure}

\section{SVCCA}
\label{appendix:svcca}
The input to SVCCA \cite{Raghu2017} are two matrices, $\bH_1 \in \mathbb{R}^{P \times N_1}$ and $\bH_2 \in \mathbb{R}^{P \times N_2}$, which hold the state vector representations of two RNNs over $P$ test inputs.
Here, $N_1$ and $N_2$ denote the number of neurons in each RNN (in general $N_1 \neq N_2$).
First, the singular value decomposition (SVD) is computed for each matrix: $\bH_1 = \bU_1 \bS_1 \bV^\top_1$ and $\bH_2 = \bU_2 \bS_2 \bV^\top_2$.
Then, these decompositions are truncated by taking the top $R$ singular vectors.
The value of $R$ is a user-defined hyperparameter.
Let $\widetilde \bV_1 \in \mathbb{R}^{R \times N_1}$ and $\widetilde \bV_2 \in \mathbb{R}^{R \times N_2}$ denote the truncated right singular vectors.
Finally, canonical correlations analysis (CCA; \cite{Hotelling1936}) is performed to quantify the similarity of $\widetilde \bV_1$ and $\widetilde \bV_2$.

\section{Centered kernel alignment (CKA)}
\label{appendix:ckaresults}
Centered kernel alignment (CKA; \cite{kornblith2019similarity}) is a measure of similarity between representations that is invariant to orthogonal transformation and isotropic scaling, but unlike SVCCA, is not invariant to invertible linear transformations. It defines a similarity between two representations $X \in \mathbb{R}^{m \times n_x}$ and $Y \in \mathbb{R}^{m \times n_y}$ where $m$ is the number of examples, and $n_x$ and $n_y$ are the number of units in the representations for $X$ and $Y$, respectively. The measure can be computed (for a linear kernel, see \cite{kornblith2019similarity} for details) as:

$$ \text{CKA}(X, Y) = \frac{ \|X^T Y \|_F^2}{\|X^T X\|_F \|Y^T Y\|_F}$$

Below, we compare representations using both singular vector canonical correlation analysis (SVCCA) and centered kernel alignment (CKA). Each figure shows comparisons for each of the three studied tasks. Within each figure, the first row shows the full pairwise distance matrix using either SVCCA (left column) or CKA (right column). The next two rows show embeddings of a subset of these pairwise distances in 2D using multi-dimensional scaling, highlighting differences by architecture (middle row) or activation (bottom row). The key takeaway is that both SVCCA and CKA show differences between network representations that cluster based on RNN architectures.

\begin{figure}[ht!]
    \centering
    \includegraphics[width=0.7\textwidth]{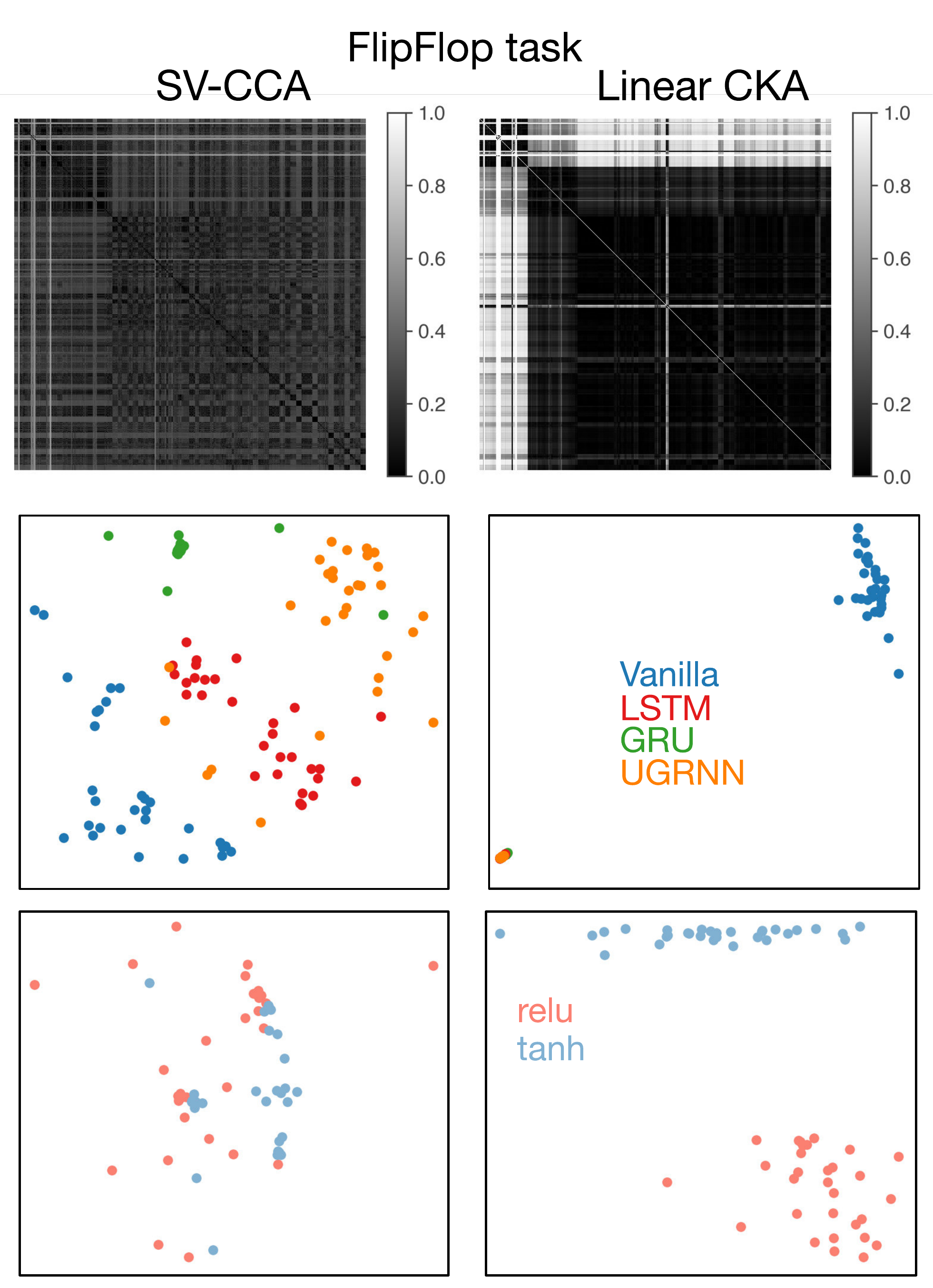}
    \caption{Comparing SVCCA and CKA for the flip flop task. See Appendix \ref{appendix:ckaresults} for description of the panels.}
    \label{fig:cka_flipflop}
\end{figure}

\begin{figure}[ht!]
    \centering
    \includegraphics[width=0.7\textwidth]{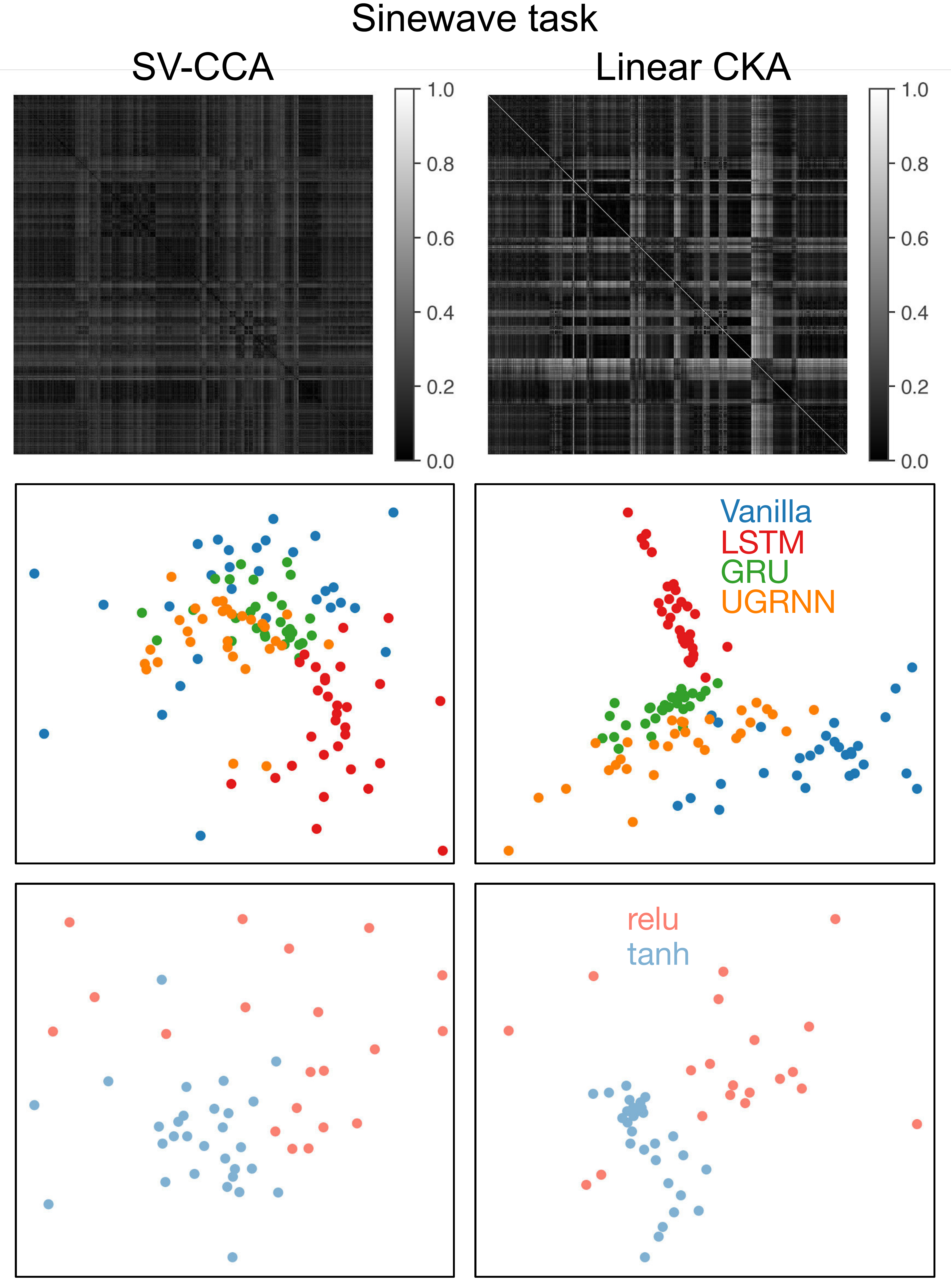}
    \caption{Comparing SVCCA and CKA for the sinewave task. See Appendix \ref{appendix:ckaresults} for description of the panels.}
    \label{fig:cka_sinewave}
\end{figure}

\begin{figure}[ht!]
    \centering
    \includegraphics[width=0.7\textwidth]{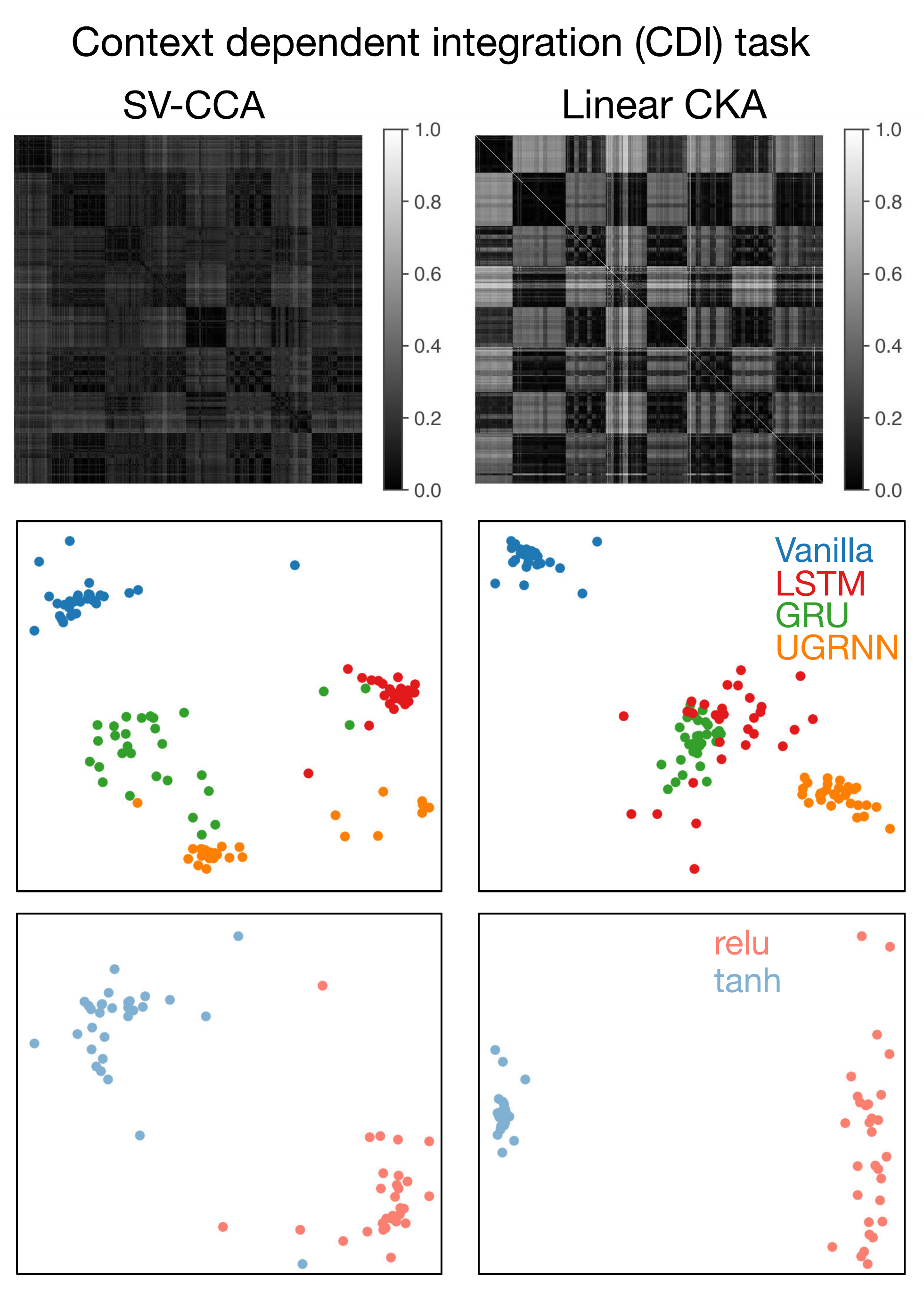}
    \caption{Comparing SVCCA and CKA for the context dependent integration task. See Appendix \ref{appendix:ckaresults} for description of the panels.}
    \label{fig:cka_cdi}
\end{figure}